# Evidence of low-dimensional chaos in magnetized plasma turbulence

*Tatjana Živkovic, Kristoffer Rypdal*
Department of physics, Auroral Observatory, University of Tromsø, Norway

**Abstract**

We analyze probe data obtained from a toroidal magnetized plasma configuration suitable for studies of low-frequency gradient-driven instabilities. These instabilities give rise to field-aligned convection rolls analogous to Rayleigh-Benard cells in neutral fluids, and may theoretically develop similar routes to chaos. When using mean-field dimension analysis, we observe low dimensionality, but this could originate from either low-dimensional chaos, periodicity or quasi-periodicity. Therefore, we apply recurrence plot analysis as well as estimation of the largest Lyapunov exponent. These analyses provide evidence of low-dimensional chaos, in agreement with theoretical predictions.

## 1. Introduction

The Helimak configuration is a plasma confined by a purely toroidal magnetic field with a weak vertical magnetic component superposed. The plasma source is an electron-emitting tungsten wire suspended vertically across the center of the torus-shaped vacuum chamber. In the experiment referred to in this paper the plasma potential and electron density fluctuations were measured by field-aligned cylindrical Langmuir probes: one of these probes was kept in a fixed position, while the other was moved in the cross section of the plasma column by a motorized, computer-guided positioning system. In every location $1 \times 10^5$ samples of electron density and potential fluctuations were measured simultaneously from both probes at the sampling frequency of 100 kHz. From the cross-coherence and cross-phase analysis of the electron density fluctuations it was shown in [1] that the dominant modes are flute-like electrostatic interchange modes, which are in a monochromatic regime when the magnetic field is below $50$ mT (close to threshold of the interchange instability) and in a turbulent regime when magnetic field is greater than $50$ mT.

It has been pointed out by several authors that the Rayleigh-Benard instability has its counterpart in flute interchange instabilities in magnetized plasmas, and a set of reduced Lorenz equations for the dynamics of plasma streamers providing cross-field plasma transport in the Helimak configuration was derived in [2]. The motivation for this paper is to investigate whether experimental data obtained from the Helimak device could reveal the existence of low dimensional, possibly chaotic, dynamics.

In a real plasma system it is not likely that one will observe "clean" low-dimensional dynamics: the typical situation is that high-dimensional turbulent dynamics dominates the smaller scales, while the low-dimensional behavior governs the global scale of the system. Moreover, experimental data are influenced by noise whose effect destroys the coherence of the dynamics on small scales. As a consequence, there is no smooth manifold containing the dynamical trajectories in any dimension of the reconstructed phase space, and standard methods such as correlation dimension analysis or false nearest neighbors method would indicate that the system is high-dimensional. On the other hand, the mean-field dimensional method, developed in [3] can be successfully applied to the experimental data. In this method the probability density function $P(D)$ is introduced. This probability density function shows us the minimum embedding dimension of the average phase space that approximates the actual dynamics. By applying mean-field dimension method to the plasma potential-electron density correlated database, we find evidence of low dimensional behavior in the Helimak data. In order to classify this low dimensionality we use recurrence plot analysis as well as Lyapunov exponent estimation.

## 2. Preparing the data

We apply a moving average filter on the original data in order to analyze the signal in the reference frame of the moving plasma, i.e. we cut off the Doppler shifted high-frequency hump in the power spectra (see Fig. 1) produced as a consequence of azimuthal rotation of the plasma column. In this way, we extract larger scale components which show the response of the background plasma which is predicted to exhibit chaotic behavior [2]. We have also applied a "mexican hat" wavelet filter in order to test the validity of the moving average filter, and we plot the filtered signal on top of the raw signal in the upper panel of Fig. 2. In the lower panel of Fig. 2 we plot the wavelet energy as a function of time *t* and frequency *f=1/a*, where *a* is the scale-parameter of the wavelet. The horizontal line drawn in the figure marks the frequency above which the wavelet coefficients are set to zero by the filter.

The filtered signal, as well as the wavelet energy spectrum, gives an impression of more or less monochromatic short bursts of frequency near 2 kHz appearing at random times, reminiscent of certain types of intermittent chaos. The analysis made in the following lends some support to this interpretation.

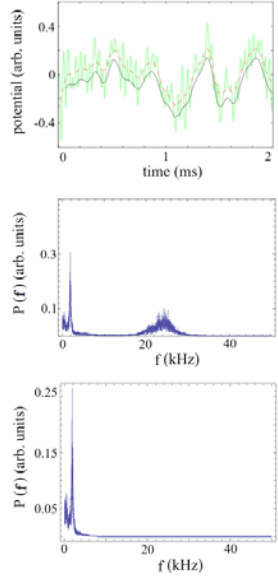

Fig. 1: Upper panel: Green line is plasma potential, dashed line is moving average, while wavelet filtered signal is in black. Middle panel: Power spectra of the plasma potential. Lower panel: Power spectra of moving average of the plasma potential signal.

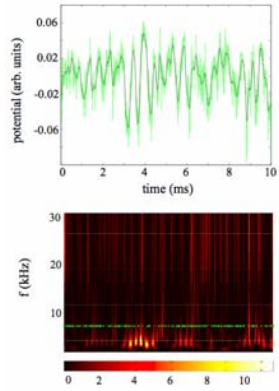

Fig. 2: Upper panel: Green line shows original time series with black wavelet filtered signal superposed. Lower panel: Wavelet energy where the range above the green line is removed by wavelet filtering. Color bar shows the intensity of the wavelet energy coefficients.

## 3. Phase space reconstruction and mean-field dimension method

In order to reconstruct the phase space, we use the time-delay embedding method [4] on the multivariate (input-output) time series and create input-output delay vectors:

$$(I_t^T, O_t^T) = (I_t, I_{t+\tau}, ..., I_{t+(m_i-1)\tau}, O_t, O_{t+\tau}, ..., O_{t+(m_o-1)\tau}), \quad (1)$$

where $\tau$ is the time delay obtained from the average mutual information [5], while $m_i$ and $m_o$ are embedding dimension for the input and output time series, respectively. Throughout our analysis, $m_i = m_o = m$. The delay matrix $M$ can be obtained from the input-output delay vectors $(I_t^T, O_t^T)$ for all times $t = 1, 2, \ldots, T - (m-1)\tau$, where T is the length of the time series. Then, the covariance matrix $C = M^T M$ can be calculated. This matrix is Hermitian and its eigenvectors form an orthonormal basis in the embedding space. We apply the singular value decomposition method to the covariance matrix in order to obtain orthonormal eigenvectors and then project the input-output delay vectors onto the space spanned by these eigenvectors. Projected input-output delay vectors are used in the following analysis and are denoted as $x_i$, where $i$ is time.

Multivariate time series are not necessary for the reconstruction of the phase space and according to the embedding theorem one time series (input time series) is sufficient. However, it has been shown in [6] that for analysis of experimental data more accurate phase space reconstruction can be obtained if data representing different observables can be applied. The low-dimensional character of the Helimak streamer dynamics can be put to a test by employing the mean-field dimensional method developed in [3]. In the following, we briefly review this method.

For a time series not contaminated by noise an essential criterion for a good embedding is that if two vectors $x_k$ and $x_n$ are close in the reconstructed embedding space, then the values of a one step iterated scalar output time series corresponding to these states $O_{k+1}$ and $O_{n+1}$ should also be close. In systems contaminated by noise, the above criterion can only be applied on vectors where noise has been reduced by averaging over N of the nearest vectors:

$$x_n^{cm} = \frac{1}{N} \sum_{k=1}^{N} x_k. \tag{2}$$

Here the superscript $cm$ is used to denote center of mass of the N nearest neighbors of the state $x_n$ in terms of Euclidean distance. The number N depends on the level of the noise in the system (for example, N is small if the noise is concentrated on the small scales).

After the nearest neighbors of each vector in the reconstructed phase space are found, one step iterated output can be approximated by:

$$O_{n+1} \approx \frac{1}{N} \sum_{k \in NN} O_{k+1}. \tag{3}$$

Here, $NN$ denotes the set of all nearest neighbors of the vector $x_n$ and $k$ denotes their indexes in the reconstructed phase space. For calculation purposes the above criterion is modified:

$$\left\|x_n^{cm} - \tilde{x}_n^{cm}\right\| \to 0 \Rightarrow \left|O_{n+1} - \tilde{O}_{n+1}\right| \to 0, \tag{4}$$

where $\tilde{x}_n^{cm}$ is the average vector which consists of all the nearest neighbors of the vector $x_n$ including $x_n$, while $x_n^{cm}$ is the average of the nearest neighbors of the vector $x_n$. Also, $\tilde{O}_{n+1}$ includes the state $x_n$ together with its nearest neighbors when calculated from Eq. 3.

The embedding is good if the inclusion of state $x_n$ in the set of its nearest neighbors does not change average output obtained from Eq. (3) significantly. We further test for which embedding dimension $D$ the conditions of Eq. (4) are satisfied for each input-output pair $(I_n, O_n)$. In other words, for which $D$ we have that $\left|O_{n+1} - \tilde{O}_{n+1}\right| \le \varepsilon$, where $\varepsilon$ is a small number which has to be chosen as a compromise between precision and sufficient statistics of our computation. When we find embedding dimensions for all the input-output pairs, the probability distribution function of the local dimension of points of the phase space whose neighbourhood fullfils Eq. (4), $P(D)$, can be computed. The typical shape of $P(D)$ is a relatively flat, or weakly humped profile for low $D$ followed by a rapid (power-law) drop to zero for larger $D$. The least embedding dimension which unfolds the attractor of the average dynamics is found to be the value of $D$ which represents the transition between the flatter core of the distribution and the rapidly decaying tail. In the case of low-dimensional dynamics, the transition from the core to the tail happens for low values of $D$ and for low level of averaging $N$. For example, $P(D)$ for the Lorenz attractor has a transition point indicating $D = 4$ for $N = 3$, in reasonable agreement with the true dimension $D = 3$. In order to see the transition in $P(D)$ for the systems with higher level of noise, one needs to do the averaging of the phase vectors for higher values of $N$. Different $N$ will give different values of $D$ and, hence, one needs to determine optimal choice of $(NN, D)$ such that the highest accuracy of the model is obtained. The accuracy can be tested by means of normalized mean squared error $(NMSE)$, where $NMSE = \frac{1}{\sigma_0}\sqrt{\frac{1}{L}\sum_{t=1}^{L}(O_t - \tilde{O}_t)^2}$. Here $t = 1$ to $L$ span the forecasting interval, $\sigma_0$ is the standard deviation of the original output time series $O_t$, while $\tilde{O}_t$ is the output time series calculated from the model.

If there were no low-dimensional attractor, no transition between a flat (or humped) core and a decaying tail for $P(D)$ could be found, indicating infinite embedding dimension. For instance, $P(D)$ for colored noise will be a power law for all $D$ and for various levels of averaging, which means that no characteristic value for $D$ exists.

Our hypothesis is that the streamer dynamics in the Helimak configuration can be described by a set of three differential equations [2] and hence the attractor dimension of its dynamics is *C≤3* (and probably *>2*). According to Takens embedding theorem [4], in the absence of noise, the embedding dimension $D$ is approximately equal to $2C+1$, and for $C \leq 3$, $D \leq 7$.

We use filtered plasma potential as an input and filtered electron density as an output in Eq. (1) and compute $P(D)$ from data obtained in various probe positions in the plasma for a wide range of values of $N$. The normalized mean square error has a minimum $NMSE \sim 0.2$ for $N= 3$, and $P(D)$ is shown in the upper panel of Fig. (3). We observe that $P(D)$ goes rapidly to zero for the embedding dimension $D>4$.

In the lower panel of Fig. 3 we plot the filtered electron density together with the one reconstructed from Eq. (3) for $N=3$ and when embedding dimension is equal to $D=4$. These results indicate low-dimensional behavior and embedding dimension $D=4$. This, however, could be the consequence of either periodic, quasi-periodic, or chaotic motion.

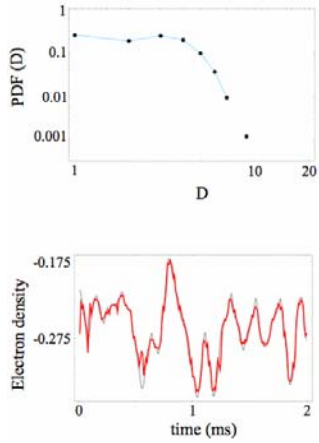

Fig. 3. Upper: Log-Log plot for the PDF (D) vs. D for N=3, Lower: Low-pass filtered electron density from the model and the actual signal.

## 4. Recurrence plot analysis

Recurrence plot analysis was introduced in [7] and has been extensively used in the last twenty years. It is easy to implement and provides a visual interpretation of the phase-space dynamics. Also, it can be applied to nonstationary data as well as to short time series, providing equally valuable results. The essential steps of the method are as follows: First the phase space is reconstructed by time-delay embedding, where vectors $x_i$ $(i=1,...,T)$ are produced. Then a $T \times T$ matrix consisting of elements 0 and 1 is constructed. The matrix element $(i, j)$ is 1 if the distance is $\|x_i - x_j\| \leq r$ in the reconstructed phase space, and otherwise it is 0. The recurrence plot is simply a plot where the points $(i, j)$ for which the corresponding matrix element is 1 are marked by a dot. The radius $r$ is fixed and chosen such that a sufficient number of points are found to reveal the fine structure of the plot. In Fig. 4, we show recurrence plots for filtered plasma potential and the electron density in the turbulent regime. We use embedding dimension $D=4$ and time delay $\tau=5$. One can easily see lines parallel with the main diagonal, which indicate recurrences of the phase-space trajectory. In the case of periodic motion, these lines are continuous and equidistant, the separation indicating the period of the motion. However, lines on Fig. 4 are relatively short and the distances between them are not equal. This can imply either the existence of quasiperiodic (several frequencies in the system, whose ratios are irrational) or chaotic motion.

The correlations between the filtered plasma potential and the electron density can be studied by means of the cross recurrence plot (CRP) and joint recurrence plot (JRP). The CRP shows dependencies between two different systems, i.e when the state of the first system recurs to the state of the other system in the same phase space [8]. On the other hand, the JRP considers recurrences of the systems in their respective phase space separately and shows the times when both of the systems recur simultaneously [9]. In Fig. 5, CRP and JRP for the filtered plasma potential and the electron density are shown. CRP reveals lines parallel with a main diagonal, which implies that filtered plasma potential and the electron density trajectories run together for some time. Also, lines in JRP show some correlations. In Fig. 6 we show JRP for the turbulent regime when the filtered plasma potential and electron density are recorded from two different positions in the plasma column. It is easy to see a presence of the recurrences. We conclude that there are correlations between the probe signals in different parts of the plasma device, demonstrating that the low-dimensional dynamics is associated with large-scale motions.

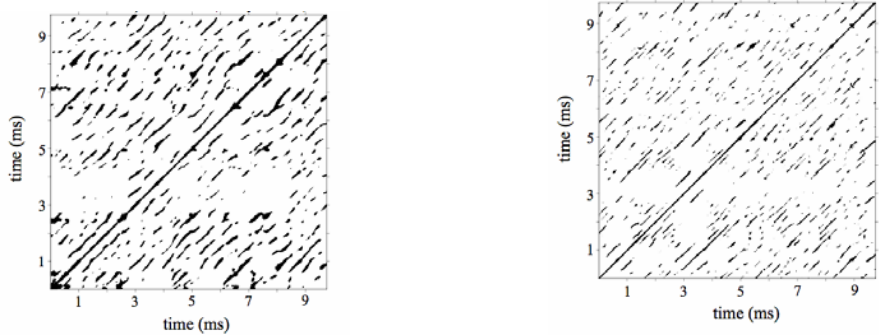

*Fig. 4. Turbulent regime. Leftr: Recurrence plot for the low-pass filtered plasma potential, Rightr: Recurrence plot for the low-pass filtered electron density*

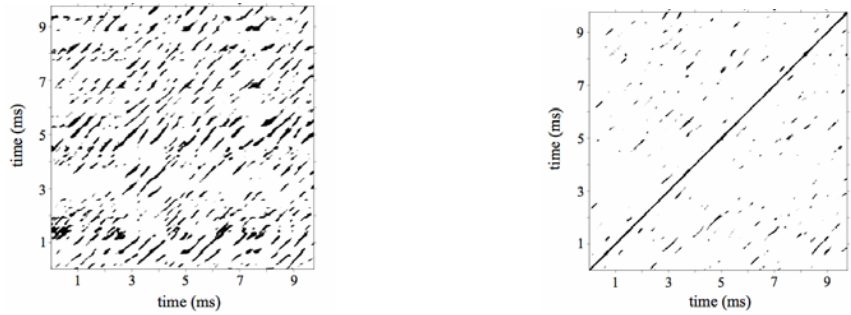

Fig. 5. Turbulent regime. Left: Cross recurrence plot for the filtered signal, Rightr: Joint recurrence plot for the filtered signal

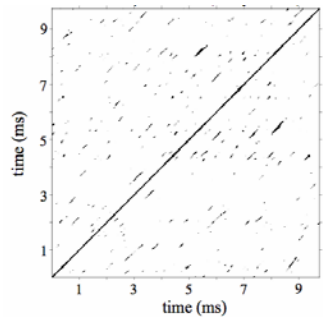

Fig. 6. Joint recurrence plot for the filtered signal when probe is in different positions.

## 5. Estimation of the Lyapunov exponent

The recurrence plot analysis provides evidence of some deterministic behavior. In order to distinguish between quasi-periodic and chaotic behavior, we need to calculate the largest Lyapunov exponent. For quasi-periodic motion trajectories do not diverge exponentially in time (are not sensitive to the initial conditions) and the largest Lyapunov exponent is not positive. To calculate the largest Lyapunov exponent we measure distances between pairs of points (fiducial point and its neighbor) and compare them after a fixed time. The logarithm of the ratio between these distances gives the current divergence of the trajectories. We then follow the evolution of the pair of points and update the Lyapunov exponent at each step, which should converge to a stationary value at the end of the analyzed time series. When the separation between points becomes large, the neighbor of the next fiducial point is chosen in such a way that the orientation of the new pair of points is as close as possible to the orientation of the original pair. Before the calculation of the Lyapunov exponent we apply a moving average to the original data. We calculate the largest Lyapunov exponent for the embedding dimension $D=4$ and time delay $\tau=5$. In Fig. 7 we plot the the largest

Lyapunov exponent as a function of time for the filtered plasma potential and density in the turbulent regime. Both of these exponents are positive and approach similar values $\lambda \approx 5$ ms$^{-1}$ , which excludes the existence of quasiperiodic motion. The Lyapunov exponent of this magnitude is consistent with the width of the power spectrum in the lower panel of Fig. 1. From Fig. 2 it is tempting to suggest that the chaotic loss of memory is associated with unpredictable intermittent onset of quasi-monochromatic pulses, but it could also be seen as unpredictable loss of phase memory in a more or less monochromatic signal.

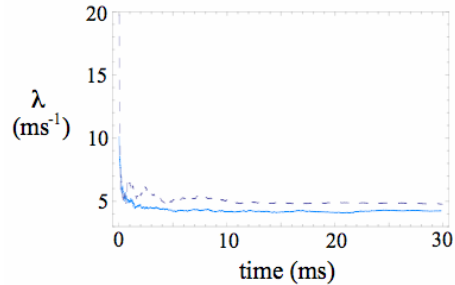

*Fig. 7. Lyapunov exponent for the low-pass filtered signal is marked as a dashed line for the electron density and is marked as a blue line for the plasma potential.*

## 6. Conclusion

We have analyzed the plasma potential and the electron density fluctuation data from a magnetically confined plasma in the Helimak configuration [1]. By applying the mean-field dimensional method [3], we conclude that the dynamics underlying these data is low-dimensional on the larger scales, and its attractor can be unfolded in the embedding dimension of $D \approx 4$. Recurrence plot analysis reveals deterministic behavior in both plasma potential and density as well as correlation between them. The largest Lyapunov exponents for the filtered plasma potential and the electron density are positive and correspond in magnitude to frequencies in the kHz range. Since these frequencies are smaller than the Doppler-shifted frequency of the observed modes in the Helimak plasma, we conclude that their origin is the oscillation of the background density and potential profile. Our analysis confirms theoretical results for plasma streamers, which were described as a system of three ordinary differential equations in [2].
These results may have relevance for other magnetized plasma configurations where interchange instability, cross-field turbulent transport, and resilient plasma profiles appear as a strongly coupled dynamic entity.

**Acknowledgements:** Cross recurrence plot and joint recurrence plot were computed by means of the Matlab package downloaded from
http://www.agnld.uni-potsdam.de/~marwan/toolbox/.